\documentclass[preprint,prb,tightenlines]{revtex4}
\usepackage{bm}
\begin{document}
\title{On boundary conditions for spin-diffusion equations 
with Rashba spin-orbit interaction}
\author{O.Bleibaum}
\email{olaf.bleibaum@physik.uni-magdeburg.de}
\affiliation{Institute for Theoretical Physics, Otto-von-Guericke University,
39016 Magdeburg, PF4120, Germany}
\begin{abstract}
We reexamine the boundary conditions of spin diffusion equations 
for dirty semiconductor heterostructures with weak linear Rashba
spin-orbit interaction. Doing so, we focus on the influence of 
tangent derivatives of the particle density at the boundary on 
the magnetization. Such derivatives are associated with a spin accumulation 
in the presence of a density gradient. We show that  the tangent 
derivatives enter the boundary conditions and argue that the
spin-Hall effect is absent in such systems because of this fact.  
\end{abstract}
\maketitle
\section{Introduction}
At present there is much interest in spin-charge coupling effects 
in non-magnetic semiconductor heterostructures with linear
Rashba spin-orbit interaction, like the spin accumulation in an external 
electric field (see, e.g., Refs.[\onlinecite{Edelstein,Kato,Silov,Inoue}]), 
spin-galvanic currents\cite{Ganichev,Ganichev2} 
or the spin-Hall effect (see, e.g., 
[\onlinecite{Sinova,Wunderlich,Nikolic,Bauer}]). While 
the first two effects can also be observed in homogeneous 
systems the spin-Hall effect manifests itself in a non-equilibrium 
magnetization at the boundaries of a sample in the presence of electric 
fields. Thus, investigations of the magnetization 
at the sample boundary require the development of a  transport theory, 
which is also able to describe the evolution of particle and spin packets.  

Problems, which are related to the motion of particle packets in 
semiconductors, have been investigated much in the literature (see,
e.g., Ref.[\onlinecite{Wallis}]). Drift-diffusion equations are 
typically used to this end. A boundary enters the diffusion equations
via boundary conditions, which have to be used in seeking solutions
to the differential equations. These boundary conditions follow often
from conservation laws, as it is the case in investigations of
charge transport processes. There the boundary conditions reduce 
to the requirement the electric current is continuous at the 
boundary. This condition implies the current vanishes at hard-wall 
boundaries. The structure of the current follows from
the conservation of the charge.

Diffusion equations for the investigation of inhomogeneous  
magnetizations in semiconductor heterostructures with 
Rashba spin-orbit interaction have been derived in a number of papers 
(see, e.g., Refs.[\onlinecite{OB1,Damker,Burkov,Mishchenko,OB2}]).
However, in most cases they have been applied to systems without 
boundaries. The problem how to formulate  boundary conditions
for systems with hard-wall boundary 
is a rather controversial one. The conventional
conservation laws can not be invoked, since the magnetization is 
not a conserved quantity in the systems in question. To circumvent this
problem some researchers have taken the point view that the 
spin current can be extracted from the derivative terms in the
diffusion equation\cite{Damker,OB2}. The
latter takes the form\cite{OB2}
\begin{eqnarray}\label{I1}
\partial_t{\bm S}\hspace{-0.5ex}&+&\hspace{-0.5ex}{\bm \Omega}\circ
({\bm S}-{\bm S}_0)+
{\bm R}\times({\bm S}-{\bm S}_0)-D\Delta{\bf S}\nonumber\\
&+&\hspace{-0.5ex}D({\bm F}\cdot
{\bm \nabla})\partial_\epsilon{\bm S}-\omega_s
({\bm N}\times{\bm \nabla})
\times({\bm S}
-{\bm S}_0)=-\partial_{\epsilon}{\bm j}_{\epsilon}
\end{eqnarray}
in the presence of a constant electric field ${\bm F}$
(${\bm S}$ is 
the spin density, $D$ is the diffusion constant,
${\bf \Omega}$ is a symmetric tensor of second rank, $\circ$ is the
dyadic product,
${\bm R}={\bm F}\times{\bm N}\partial_{\epsilon}\omega_s/2$ is a vector,
$\omega_s=4mD/\hbar$ is a transport coefficient,
 ${\bm N}=N{\bm e}_z$ is a vector 
perpendicular to the  plane of the two-dimensional electron gas, 
$\partial_{\epsilon}$ is the derivative with respect to energy, 
${\bm j}_{\epsilon}$ is the energy current for spins and 
${\bm S}_0=-\tau{\bm N}\times{\bm\nabla} n+\tau {\bm N}\times{\bm F}
\partial_{\epsilon}n$ is the spin accumulation, which can be induced either by 
the 
constant field  ${\bm F}$ or by a gradient of the density $n$). 
A characteristic feature
of this equation is the fact most of its terms depend only on the 
deviation of the magnetization from the spin accumulation ${\bm S}_0$.
Thus, the point view that all derivative terms are part of the divergence
of the spin-current tensor leads  to the conclusion that the derivative
of the magnetization normal to the boundary is proportional to 
${\bf S}-{\bf S}_0$ 
and thus
zero in the stationary state. Consequently, there is no spin-Hall effect
in such systems in the presence of a hard-wall boundary. This point of view 
is supported further by the fact that no anomalous Hall-current is induced by
injection of particles with perpendicular spin in 
electric fields\cite{OB2}.

However, although diffusion currents have been extracted from 
diffusion equations for centuries it has to be said that 
this approach is problematic.  The diffusion equations contain only 
the divergence of the spin-current tensor, so terms proportional
to a curl can not be retrieved unambiguously by this procedure.
Moreover, the interpretation of  terms as part of the divergence
of a tensor of second rank is also ambiguous. The sixth term on the
left hand side (lhs) of Eq.(\ref{I1}), e.g., can also be decomposed into 
the gradient of a scalar field and the curl of a vector field.  This raises 
the question, whether there is another procedure, which can be used to
determine boundary conditions for  diffusion equations. 

Recently, such a method has been developed in Ref.[\onlinecite{Galitski}]. 
Starting from the notion, that the boundary conditions for the system in
question can be expressed in the form 
\begin{equation}\label{I2}
{\bm e}_n\cdot{\bm \nabla}\rho_{\alpha}=B_{\alpha\beta}\rho_{\beta},
\end{equation}
which connects the normal derivative ${\bm e}_n\cdot {\bm\nabla}\rho_{\alpha}$
of the densities $\rho_{\alpha}=(n,{\bm S})$ with the densities themselves,
they calculate the coefficients $B_{\alpha\beta}$ from the integral 
equations for the densities (A 
summation with respect to double indices must be performed in all 
equations.). Doing so, they use the fact that the integrals are determined 
by two scales near the boundary, by the Fermi wave-length and the
mean free path. 
The first length is the distance from the boundary needed to approach a 
constant particle density and the second the distance needed to feel the 
impact of disorder. Thus, transport is quasi-ballistic in the 
intermediate region. After a tour the force through rather complicated 
integrals they find that the coefficients coupling spin and charge vanish
in the intermediate region and that those coupling spin and spin are 
non-zero. This fact has a very
important consequence. The derivatives of the
magnetization do not vanish at the boundary in this case, since the 
magnetization tends to approach the value ${\bm S}_0$ in the stationary limit 
(see Eq.(\ref{I1})). Therefore, spin-Hall effect is obtained.

The method developed in Ref.[\onlinecite{Galitski}] has proven to be 
very useful in the investigation of the boundary conditions. 
The results of Ref.[\onlinecite{Galitski}], however,
are neither in line with those of previous investigations of the structure of 
the diffusion equations\cite{OB2}
nor with the investigations of the spin-current tensor. Already a glance 
on the spin-diffusion equations
(\ref{I1}) reveals  we should expect that the boundary conditions
are not determined by the magnitude of the magnetization but by the deviation
of the magnetization from the spin accumulation ${\bf S}_0$. The spin 
accumulation, however, is determined by tangent derivatives. Such derivatives
are not contained in Eq.(\ref{I2}).  Thus, the 
ansatz (\ref{I2}) is simply not sufficiently general enough to accommodate 
the spin accumulation. To allow for an impact of the spin accumulation we
should replace  the ansatz (\ref{I2})  by the new ansatz
\begin{equation}\label{I3}
{\bm e}_n\cdot{\bm \nabla}\rho_{\alpha}=B_{\alpha\beta}\rho_{\beta}
+A_{\alpha\beta}\;\; {\bm e}_t\cdot{\bm\nabla}\rho_{\beta}
\end{equation}
and determine the spin-charge coupling coefficients of the matrix 
$A_{\alpha\beta}$ (${\bm e}_t$ is the unit tangent vector). 
Boundary conditions of the type (\ref{I3}) are not unusual
for partial differential equations. Indeed, also the boundary 
conditions for Maxwells equations can be formulated in 
terms of normal and tangent derivatives of fields. 

Motivated by our observation we now calculate the spin-charge coupling 
coefficients of the matrix $A_{\alpha\beta}$ for systems with 
weak Rashba interaction in the absence of external fields.
To this end we focus on the situation in a half space and assume that 
there is a gradient in the charge carrier density tangent
to the boundary.  Moreover, we assume that the normal derivative
of the charge carrier density vanishes. This is just the situation
in a spin-Hall experiment. The spin-spin coupling 
coefficients are of higher order with respect to the Rashba interaction
and therefore negligible in the systems in question. 
\section{Basic equations}
The Hamilton operator for the two-dimensional electron gas has the form
\begin{equation}\label{H1}
H=\frac{{\bm{\hat p}}^2}{2m}+{\bm N}\cdot({\bm \sigma}\times{\bm{\hat p}})
+V({\bm { x}}).
\end{equation}
Here ${\bm{\hat p}}$ is the momentum operator, ${\bm\sigma}$ is the 
spin-operator and $V({\bm { x}})$ is a random potential with zero average,
Gaussian statistics and mean squared deviation
\begin{equation}\label{H2}
\langle V({\bm x})V({\bm x'})\rangle =\frac{\hbar}{2\pi\nu\tau}
\delta({\bm x}-{\bm x'}).
\end{equation}
The unit vectors in the 2-d plane are ${\bm e}_x$ and ${\bm e}_y$. 
${\bm e}_z$ is a unit vector perpendicular to the plane. 
To be specific we consider the half-space $y>0$. Thus, derivatives with 
respect to $y$ are normal to the boundary and derivatives with respect to
$x$ are tangent to the boundary.
The labels $x$,$y$ and $z$ are replaced 
by the indices 1,2,3 in sums. Latin indices run from 1-3, Greek indices from
0-3. The index zero is associated with the particle density and $\sigma_0$ is
the unit matrix. Double indices have to be summed over in 
all equations.  

The propagation of single particle excitations is described by the 
retarded  
and advanced Green's functions. To calculate them we use
the fact  that the system is translation invariant in 
$x$-direction in average, so we we can use the Fourier transformation. 
Doing so, we find that the retarded (R) Green's functions is given by 
\begin{equation}\label{H3}
G_0^{R}(y,y'|k)=\frac{m}{\lambda_{R}\hbar^2}(
e^{-\lambda_{R}(y+y')}-e^{-\lambda_{R}|y-y'|})
\end{equation} 
in the absence of the spin-orbit interaction, where
\begin{equation}\label{H4}
\lambda_{R}=- i\sqrt{ k_F^2-k^2+ \frac{ik_F}{l}}.
\end{equation}
Here $\hbar k_F$ is the momentum at the Fermi surface and $l$ is the mean 
free path. The advanced (A) Green's function is obtained from Eq.(\ref{H3})
by hermitian conjugation.
To take into account the spin-orbit interaction we use the Dyson equation 
approach of Ref.[\onlinecite{Galitski}]. The spin-orbit
interaction is considered as the perturbation in this approach 
and Eq.(\ref{H3}) as the 
leading approximation. Thus, the approach ignores the impact of the boundary 
on the scattering time. The first correction with respect to the spin-orbit 
interaction is given by $G_1^{R}=\delta_1 G^{R}_1+\delta_2 G^{R}_1$, where
\begin{equation}\label{H5}
\delta_1 G^{R}_1(y,y'|k)=-\sigma_y
\frac{Nkm^2}{\lambda^2_{R}\hbar^3}[e^{-\lambda_{R} |y-y'|}
(|y-y'|+\frac{1}{\lambda_{R}})-e^{-\lambda_{R}(y+y')}
(y+y'+\frac{1}{\lambda_{R}})]
\end{equation}
and
\begin{equation}\label{H6}
\delta_2G^{R}_1(y,y'|k)=\sigma_x\frac{im^2N}{\lambda_{R}\hbar^3}
(y-y')[e^{-\lambda_{R} |y-y'|}-e^{-\lambda_{R}(y+y')}].
\end{equation}
This correction is sufficient
to calculate the matrix $B_{\alpha\beta}$ for systems with weak
Rashba interaction. Therefore, the consideration has been restricted to 
this contribution in previous calculations\cite{Galitski}. 
The investigation of 
spin-charge coupling effects, however, requires to go beyond this 
approximation. The leading contributions to these coefficients result from
corrections of second order with respect to the spin-orbit interaction
odd in $k$ in the situation discussed here. 
Therefore, we also need the odd part of the second order
correction to the Green's function. It is given by
\begin{equation}\label{H7}
G_2^R(y,y'|k)|_{odd}=\frac{2N^2km^3}{\lambda_R^2\hbar^4}yy'\sigma_z
e^{-\lambda_R(y+y')}.
\end{equation}
%
 
\section{The boundary conditions}
To extract the boundary conditions we use the fact
that the densities satisfy the system of integral equations
\begin{equation}\label{B1}
\rho_{\alpha}({\bm x})=\frac{\hbar}{4\pi\nu\tau}
\int d{\bm x}_1
\mbox{tr}
(\sigma_{\alpha}G^R({\bm x},{\bm x}_1)\sigma_{\beta}
G^A({\bm x}_1,{\bm x}))\rho_{\beta}({\bm x_1})
\end{equation}
in the stationary limit. Here the Green's functions are those 
in position representation. To simplify this expression we expand the 
integrals at the point ${\bm x}$ and obtain
\begin{eqnarray}\label{B2}
\rho_{\alpha}({\bm x})&=&\rho_{\beta}({\bm x})\frac{\hbar}{4\pi\nu\tau}
\int\limits_0^{\infty}dy_1\int\frac{dk}{2\pi} 
\mbox{tr}
(\sigma_{\alpha}G^R(y,y_1|k)\sigma_{\beta}
G^A(y_1,y|k))\nonumber\\
& &+\partial_y\rho_{\beta}({\bm x})\frac{\hbar}{4\pi\nu\tau}
\int\limits_0^{\infty}dy_1\int\frac{dk}{2\pi} 
(y_1-y)\mbox{tr}
(\sigma_{\alpha}G^R(y,y_1|k)\sigma_{\beta}
G^A(y_1,y|k))\nonumber\\
& &+\partial_x\rho_{\beta}({\bm x})\frac{\hbar}{4\pi\nu\tau}
\int\limits_0^{\infty}
dy_1\int\frac{dk}{2\pi} 
\mbox{tr}
(\sigma_{\alpha}G^R(y,y_1|k)\sigma_{\beta}
i\partial_k G^A(y_1,y|k)).
\end{eqnarray}
The first two terms on the right hand side (rhs) of Eq.(\ref{B2})
are those considered in Ref.[\onlinecite{Galitski}]. The third
term is the new term associated with the tangent derivative of 
the fields. To calculate the integrals we choose a point ${\bm x}$, 
which satisfies the conditions $k_F y\gg 1$ and $y/l\ll 1$.
Doing so, we find that Eq.(\ref{B2}) can be written in the 
form
\begin{equation}\label{B3}
0=-D\partial_y S_x,\hspace{2em}0=-D\partial_yS_y-\frac{1}{2}N
\omega_sS_z,
\end{equation}
\begin{equation}\label{B4}
0=-D\partial_yS_z+\frac{1}{2}N\omega_s(S_y-{S_0}_y),
\end{equation}
where $D$ is the diffusion coefficient and ${S_0}_y$
is the spin accumulation induced by the density gradient. We would like to 
mention that the terms proportional to the magnetization in the 
Eqs.(\ref{B3}) and (\ref{B4}) agree with those discussed in 
Ref.[\onlinecite{Galitski}]. The second term in the bracket of Eq.(\ref{B4}),
however, is a new term, which couples spin and charge. Eq.(\ref{B4}) shows, 
that
the normal derivative of the magnetization depends only on the deviation
of the magnetization from the spin accumulation ${\bm S}_0$. Thus, there 
is no spin-Hall effect in systems with hard-wall boundary, as 
detailed in Ref.[\onlinecite{OB2}].

We  would like to mention that the use of the Fourier-transformed Green's 
functions simplifies the integrals strongly. If we use  the 
Eqs.(\ref{H3})-(\ref{H7})
we obtain a number of integrals, which contribute to Eq.(\ref{B2}). However, 
none of them is as complicated as those considered in 
Ref.[\onlinecite{Galitski}], where Green's functions in position space were 
used. The integrals with respect to $y_1$ reduce to simple exponential 
integrals, which can be calculated immediately. The remaining integrands
reduce either to functions of the type $f(k)$, to functions of the
type $y^{\alpha}
f(k)\exp(-\mbox{Re}\lambda_R y)$ or to functions of the  type 
$y^{\alpha}\mbox{Re}(f(k)\exp(-\lambda_R y))$ (here $f(k)$ are complex
functions, which do not cause problems). The latter are those, which are 
strongly oscillating, as discussed in Ref.[\onlinecite{Galitski}].
All of the integrals simplify strongly
if the chain of substitutions $k=k_F x$, $x^2-1= \epsilon \sinh(z)$, 
$\exp(-z/2)=w$ is used ($\epsilon=1/k_Fl$). The intermediate scale
$k_F^{-1}\ll y\ll l$ can immediately be recognized after these substitutions
and a complete asymptotic expansion can be obtained with methods as simple
as parts integration. We would like to mention, that there is also the new 
parameter $\sqrt{\epsilon}k_Fy$ in the integrals. 
The leading contributions, however, do not depend on this parameter.
\section{On the structure of the diffusion equation}
If we compare the boundary conditions (\ref{B3}) and (\ref{B4}) with
those of Ref.[\onlinecite{OB2}] we note
they differ from those of 
Ref.[\onlinecite{OB2}] by a factor 1/2. Although this fact does not affect 
the conclusions of Ref.[\onlinecite{OB2}] there is the question what the 
source of this factor is. A glance at Eq.(\ref{I1}) reveals, that the
factor 1/2 implies, that only 1/2 of the sixth term on the lhs of 
Eq.(\ref{I1}) is a part of the divergence of the spin-current tensor.
Thus, we can write Eq.(\ref{I1}) in the form
\begin{equation}\label{D1}
\partial_t{\bm S}+{\bm\Omega}\circ({\bm S}-{\bm S}_0)+
{\bm{\tilde R}}\times
({\bm S}-{\bm S}_0)+\mbox{Div}J=-\partial_{\epsilon}{\bm j}_{\epsilon},
\end{equation}
where
\begin{equation}\label{D2}
\mbox{Div}J=-D\Delta{\bm S}+D({\bm F}\cdot{\bm\nabla})
\partial_{\epsilon}{\bm S}
-\frac{1}{2}\omega_s({\bm N}\times{\bm \nabla})
\times({\bm S}-{\bm S}_0),
\end{equation}
and 
\begin{equation}\label{D3}
{\bm{\tilde R}}={\bm R}-\frac{1}{2}\omega_s{\bm N}\times{\bm\nabla}.
\end{equation}
Eq.(\ref{D1}) differs from conventional diffusion equations for the
magnetization only in that the third term on the rhs is proportional
to ${\bm S}-{\bm S}_0$. The term ${\bm{\tilde{R}}}\times{\bm S}_0$ 
is absent in conventional
diffusion equations, since ${\bm{\tilde {R}}}||{\bm S}_0$ in them.
Thus, the presence of this term in the diffusion equation gives rise to the 
notion that
the impact of the  field differs from that of a density gradient, 
as already discussed in Ref.[\onlinecite{OB2}]. However,
if we take into account the structure of the vectors ${\bm R}$  
and ${\bm{\tilde{R}}}$ we can also write this 
equation in the form
\begin{equation}\label{D4}
\partial_t{\bm S}+{\bm\Omega}\circ({\bm S}-{\bm S}_0)+{\bm{\tilde R}}
\times {\bm S}+\mbox{Div}J=-\partial_{\epsilon}({\bm j}_{\epsilon}
+\frac{1}{2}\omega_s({\bm N}\times{\bm\nabla})\times({\bm N}\times{\bm F})n).
\end{equation}
This fact shows that the impact of the field on the system in question
is the same as the impact of a density gradient, at least on the level 
of the diffusion equations. Thus, the fact that only 1/2 of the sixth term
enters the spin-current tensor restores the symmetry between the response
to an electric field and the response to a density gradient.
Eq.(\ref{D4}) also indicates that the
second term on the rhs of Eq.(\ref{D4}) is a part of the energy relaxation 
current. The diffusion equations for systems with Rashba spin-orbit 
interaction reduce just to the conventional diffusion equations in 
this case.
 
We stress that the conjecture that the second term on the rhs of 
Eq.(\ref{D4}) is part of the energy current still needs justification.
The present paper clarifies the boundary conditions in position space.
The energy space is a different subject, which we hope to address
in a future publication. 
\section{Conclusions}
In the present paper we have reinvestigated the boundary conditions
for spin diffusion equations with Rashba spin-orbit interaction. 
Our investigation shows that there are also spin-charge coupling terms
in the boundary conditions. They couple the tangent derivative of the 
particle density to the normal derivative of th spin-density. There is no 
spin-Hall effect in systems with weak Rashba spin-orbit
interaction in the presence of a hard-wall boundary due to the
spin-charge coupling terms. 
\begin{acknowledgments}
The author is grateful to B. K. Nikoli\'c for drawing his attention to 
Ref.[\onlinecite{Galitski}] and to H. B\"ottger for many useful
and stimulating discussions.
\end{acknowledgments}


\begin{thebibliography}{40}
\bibitem{Edelstein} V. M. Edelstein, Solid Sate Commun. {\bf 73}, 233 (1990).
\bibitem{Kato} Y. K. Kato, R. C. Myers, A. C. Gossard, and D. D. Awschalom, 
Phys. Rev. Lett.{\bf 93}, 176601 (2004).
\bibitem{Silov} A. Yu. Silov, P. A. Blajov, J. H. Woller, R. Hey, K. H. Ploog, 
and N. S. Averkiev, Appl. Phys. Lett. {\bf 85}, 5929 (2004).
\bibitem{Inoue} J. I. Inoue, G. E. W. Bauer, and L. W. Molenkamp, Phys. Rev. 
B {\bf 67}, 033104 (2003). 
\bibitem{Ganichev} S. D. Ganichev, E. L. Ivchenko, S. N. Danilov, J. Eroms, W. Wegscheider, D. Weiss, W. Prettl, Phys. Rev. Lett. {\bf 86}, 4358 (2001).
\bibitem{Ganichev2} S. D. Ganichev, V. V. Belkov, L. E. Golub, E. L. Ivchenko, P. Schneider, S. Giglberger, J. Eroms, J. DeBoeck, G. Borghs, W. Wegscheider,
D. Weiss, and W. Prettl, Phys. Rev. Lett. {\bf 92}, 256601 (2004).
\bibitem{Sinova} J. Sinova, D. Culcer, Q. Niu, N. A. Sinitsyn, T. Jungwirth, and A. H. MacDonald, Phys. Rev. Lett.{\bf 92}, 126603 (2004).
\bibitem{Wunderlich} J. Wunderlich, B. Kaestner, J. Sinova, T. Jungwirth, Phys. Rev. Lett. {\bf 94}, 047204 (2005).
\bibitem{Nikolic} B. K. Nikoli\'c, L. P. Zab\^o, and S. Souma, Phys. Rev. B {\bf 72}, 075361 (2005), Phys. Rev. B {\bf 73}, 075303 (2006).
\bibitem{Bauer} I. Adagideli and G. E. W. Bauer, Phys. Rev. Lett. {\bf 95}, 256602 (2005).
\bibitem{Wallis} M. Balkanski and R. F. Wallis, {\it Semiconductor 
Physics and Applications}, Oxford University Press (2000).
\bibitem{OB1} O. Bleibaum, Phys. Rev. B {\bf 69}, 205202 (2004).
\bibitem{Damker} T. Damker, H. B\"ottger, V. V. Bryksin, Phys. Rev. B {\bf 69},
205327 (2004).
\bibitem{Burkov} A. A. Burkov, A. S. Nunez, and A. H. MacDonald, Phys. Rev. 
B {\bf 70}, 155308 (2004).
\bibitem{Mishchenko} E. G. Mishchenko, A. V. Shytov, and B. I. Halperin, 
Phys. Rev. Lett. {\bf 93}, 226602 (2004).
\bibitem{OB2} O. Bleibaum, Phys. Rev. B {\bf 73}, 35322 (2006), 
phys. stat. sol. (b) {\bf 243}, 403 (2006).  
\bibitem{Galitski} V. M. Galitski, A. A. Burkov and S. Das Sarma, 
cond-mat/0601677v2 (2006).
\end{thebibliography}
\end{document}